\documentclass[letter,twocolumn]{jpsj3}
\usepackage{txfonts}
\usepackage{bm}

\newcount\pdf
\ifnum \pdf=1
\else
\fi

\title{Line-Node Dirac Semimetal 
and Topological Insulating Phase 
in Noncentrosymmetric Pnictides CaAg$X$ ($X= \mathrm{P}, \mathrm{ As}$)}

\author{Ai Yamakage,$^{1,3}$ Youichi Yamakawa,$^{2,3}$ Yukio Tanaka,$^1$ and Yoshihiko Okamoto$^{1,3}$}
\inst{$^1$Department of Applied Physics, Nagoya University, Nagoya 464-8603, Japan \\
$^2$Department of Physics, Nagoya University, Nagoya 464-8602, Japan \\
$^3$Institute for Advanced Research, Nagoya University, Nagoya 464-8601, Japan}

\abst{Two noncentrosymmetric ternary pnictides, CaAgP and CaAgAs, are reported as  topological line-node semimetals protected solely by mirror-reflection symmetry. The band gap vanishes on a circle in momentum space, and surface states emerge within the circle. Extending this study to spin-orbit coupled systems reveals that, compared with CaAgP, a substantial band gap is induced in CaAgAs by large spin-orbit interaction. The resulting states are a topological insulator, in which the $\mathbb Z_2$ topological invariant is given by $1;000$. To clarify the $\mathbb Z_2$ topological invariants for time-reversal-invariant systems without spatial-inversion symmetry, we introduce an alternative way to calculate the invariants characterizing a line node and topological insulator for mirror-reflection-invariant systems.}

\makeatletter
\AtBeginDocument{\@ifpackageloaded{natbib}{\ifNAT@numbers\if@filesw\immediate\write\@auxout{\string\global\string\NAT@numberstrue}\fi\fi}{}}
\makeatother
\begin{document}

\maketitle


\textbf{\textit{Introduction.--}}
The idea of topology has been greatly expanded in the field of condensed matter physics.
The quantum Hall effect can be viewed as a topological insulating state.\cite{thouless82, kohmoto85}
In the past decade, various topological insulators respecting time-reversal symmetry  have been discovered and this has motivated numerous subsequent studies.\cite{hasan10, qi11, tanaka12}
However, zero-gap semiconductors (Weyl/Dirac semimetals)\cite{murakami07, wan11, young12, steinberg14}
have recently been  recognized as a topologically nontrivial system.
The absence of a band gap and the approximately linear dispersion in the low-energy regime are the characteristic features of the Weyl/Dirac semimetals.
Because the low-lying excitations are the same as those of relativistic massless fermions, i.e., Weyl/Dirac fermions, anomalous transport phenomena such as the chiral magnetic effect\cite{fukushima08} can be expected not only in high-energy physics but also in solids.\cite{zyuzin12, hosur13}
Thus, exploring topological semimetals can serve as a basis for understanding novel phenomena in condensed matter physics.

In  most Weyl/Dirac semimetals,
conduction bands overlap with valence bands at certain momentum points. 
However, the band gap rarely vanishes on a momentum line.\cite{burkov11, chiu14, fang15, gao}
Such a dispersion structure is called a ``line node," which is analogous to that in  line-node superconductors. 
Recently, many systems have been proposed as  line-node semimetals (e.g., graphite\cite{mikitik06, heikkila11, heikkila}, the heterostructure of topological insulators,\cite{phillips14} hyperhoneycomb lattice,\cite{mullen15} transition-metal monophosphides,\cite{weng-bernevig} carbon allotropes,\cite{weng15, chen} Cu$_3$N\cite{kim-kane}, antiperovskites,\cite{yu15} rare-earth monopnictides,\cite{zeng} and perovskite iridates\cite{carter12, chen15, kim15, liu}). 
Furthermore, several quantum phenomena are also expected to appear in  line-node semimetals; these include a flat Landau level,\cite{rhim15} long-range Coulomb interaction,\cite{huh2} the Kondo effect\cite{mitchell}, and a quasi-topological electromagnetic response that induces charge polarization and orbital magnetization proportional to the length of the line node.\cite{ramamurthy} 
Experimental results on line-node semimetal materials are also being reported.\cite{xie15, bian, schoop}
Interestingly, in addition to these materials, photonic crystals\cite{lu12} and spin liquids\cite{natori} have been shown to host  line nodes.
The line-node structure of energy bands is becoming a prominent topic.

\begin{figure*}[ht]
\centering
\ifnum \pdf=0
\includegraphics{fig1v2+_outline.eps}
\else
\includegraphics{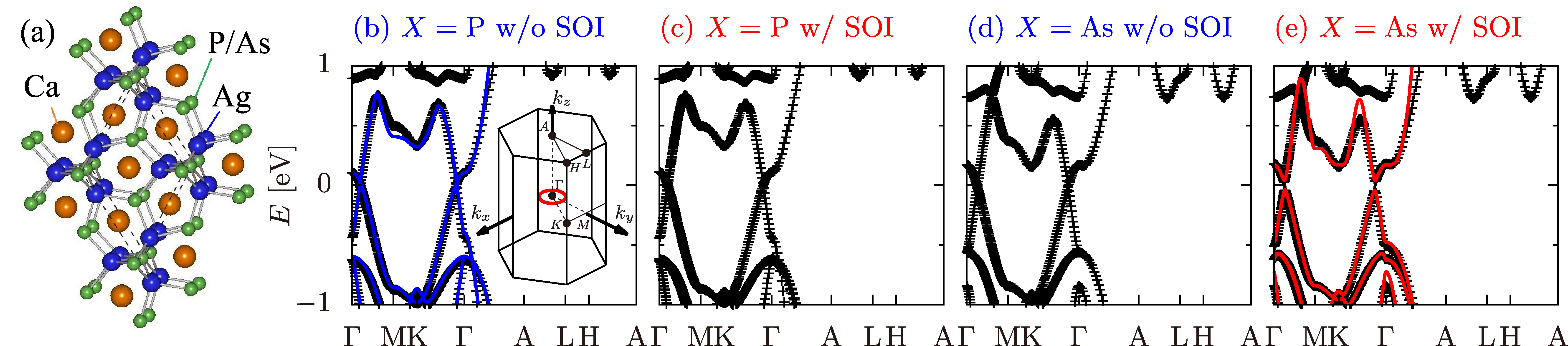}
\fi
\caption{(a) Crystal structure of CaAg$X$ ($X = \mathrm P$, $\rm As$). 
Small, middle, and large spheres represent $X$, Ag, and Ca atoms, respectively.
The dotted lines show the hexagonal unit cell. 
(b)--(e) Electronic states of CaAgP and CaAgAs with (w/) and without (w/o) spin-orbit interaction (SOI).
The crosses in (b)--(e) and the solid curves in (b) and (e) indicate the band dispersion obtained by first-principles calculations and by using the tight-binding models, respectively.
In the inset of (b), the location of the line node is indicated by the (red) circle around the $\Gamma$ point in the Brillouin zone.}
\label{fig1}
\end{figure*}

In this Letter, we propose hexagonal pnictides CaAg$X$ ($X = \mathrm P$ and As) as  novel line-node Dirac semimetals. 
Mewis synthesized these compounds and found that they crystallize in the ZrNiAl-type structure with space group P$\bar 6$2m.\cite{mewis79}
As depicted in Fig. \ref{fig1}(a), Ag$X_4$ tetrahedra form a three-dimensional network by sharing their edges and corners with intervening Ca atoms, which form a kagome-triangular lattice.\cite{ishikawa14} 
An important aspect of this structure in terms of the physics of topology is that the space group has  D$_{\rm 3h}$ point-group symmetry, 
i.e., mirror-reflection symmetry is preserved while spatial-inversion symmetry is not, although
almost all of the previously studied line-node semimetal materials have spatial-inversion symmetry.
First-principles calculations and symmetry  consideration indicate that there actually exists a line-node in CaAg$X$. 
Additionally, the $\mathbb Z_2$ topological invariant regarding a line node is well defined by using  mirror-reflection symmetry instead of spatial-inversion symmetry. 
The relation between the topological invariant and surface states is also discussed.
When the spin-orbit interaction is switched on,
the line node disappears and the system turns into a topological insulator. 
We investigate the topological phase of CaAg$X$ both from the energy dispersion on the surface and from the $\mathbb Z_2$ topological invariants, which is obtained to be $\nu_0;\nu_1 \nu_2 \nu_3 = 1;000$.



\textbf{\textit{Bulk electronic states.--}}
First, we perform the first-principles calculations
 by using the WIEN2k code.\cite{WIEN2k, DFT}
We use the experimental structural parameters\cite{mewis79} for the calculations.
Figure 1(b) shows the calculated band structures of CaAgP without the spin-orbit interaction. 
The line node is observed within 10 K from the Fermi level along the $\Gamma M$ and $\Gamma K$ lines, 
and it forms a circle in the $k_x$-$k_y$ plane centered at the $\Gamma$ point, as illustrated in the inset of Fig. \ref{fig1}(b).
The band dispersion at the line node is linear along both the radial and $k_z$ directions. 
This line node is not protected for the spin-orbit interaction. 
However, as shown in Fig. 1(c), the effect of the spin-orbit interaction in CaAgP is negligible and the size of the induced gap at the line node is of the order of $10$~K because of the weak spin-orbit interaction in the P atom. 
In stark contrast, the spin-orbit interaction has a significant effect on the band structures of CaAgAs. 
We show the calculated band structure of CaAgAs without and with spin-orbit interaction
 in Figs. 1(d) and 1(e), respectively. 
The line node has a large gap of $\sim$1000~K, indicating that the bulk system is an insulator.

Next, we derive tight-binding models for CaAgP and CaAgAs to investigate the surface electronic states. 
According to  first-principles calculations,
the main component of the conduction band at the $\Gamma$ point is the $p_z$ ($A''_2$ in $\rm D_{3h}$) orbital of P or As atoms. 
In contrast, the valence band around the $\Gamma$ point mainly consists of the $s$ orbital ($A_1'$ in $\rm D_{3h}$) of Ag atoms in addition to the $p_x$ and $p_y$ orbitals ($E'$ in $\rm D_{3h}$) of P or As atoms. 
Therefore, we construct the 12-orbital tight-binding models by constructing the maximally localized Wannier functions for the $3p$ ($4p$) orbitals of three P (As) atoms and the $5s$ orbital of three Ag atoms in CaAgP (CaAgAs).\cite{marzari97, souza01}
Furthermore, we have checked that the results do not alter even if the $d$ orbitals of Ca atoms are taken into account.\cite{dorbital}
Here, the spin-orbit interaction $H_{\rm SO} = \lambda {\bm L} \cdot {\bm S},$
 with $\lambda = 0.07$ eV for $4p$ electrons, is taken into account in the CaAgAs model,
 whereas it is neglected in the CaAgP model. 
In Figs. 1(b) and 1(e), we see a good agreement between the first principle band and the obtained tight-binding band.

\begin{figure}
\centering
\ifnum \pdf=1
\includegraphics{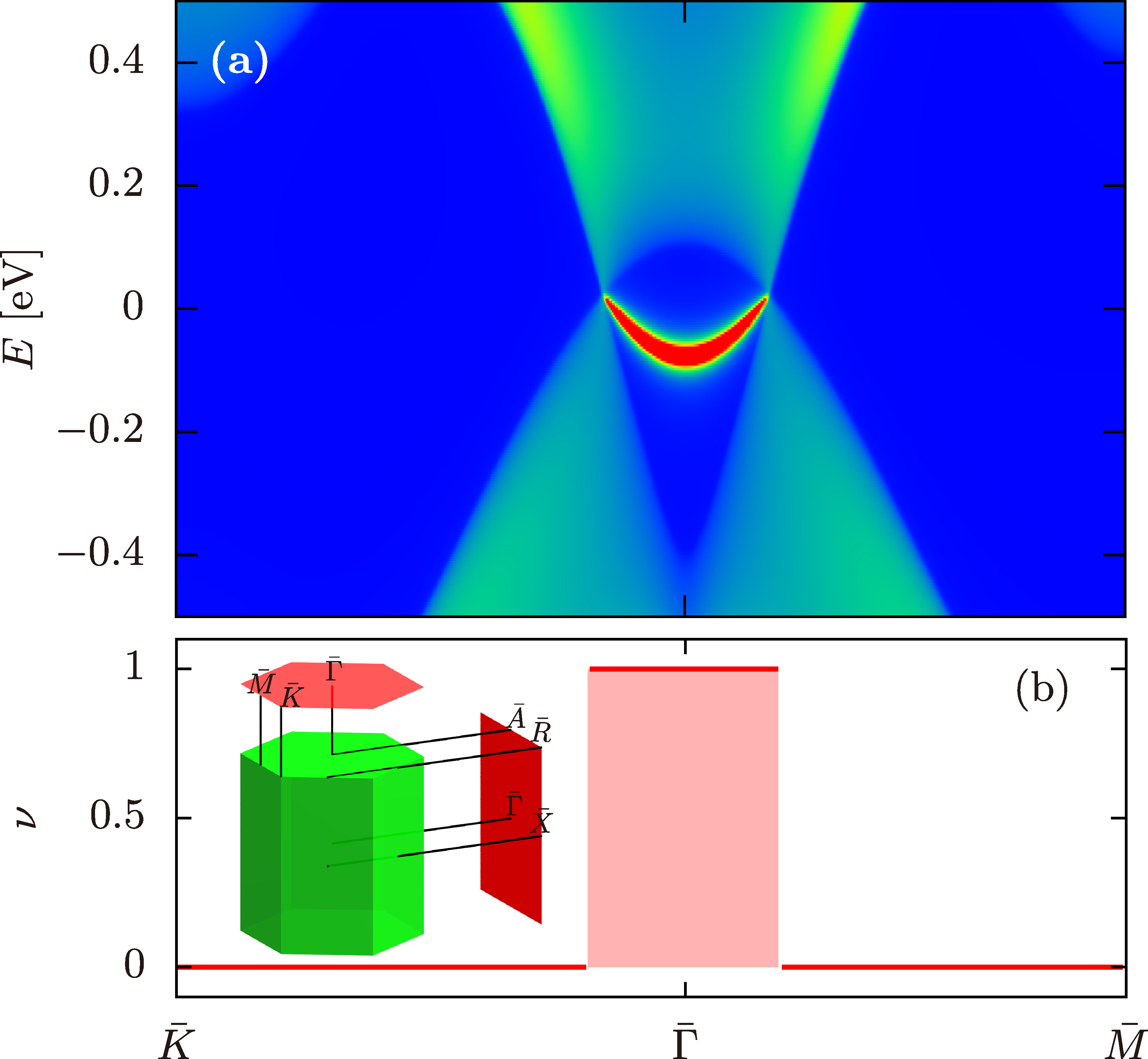}
\else
\includegraphics{fig2v2_outline.eps}
\fi
\caption{Angle-resolved density of states on the (0001) surface with the Ca$_3 \mathrm P$ termination (a) and the topological number $\nu$ of CaAgP (b). The inset in (b) shows the projected Brillouin zones onto the (0001) and $(10 \bar 10)$ surfaces.}
\label{fig2}
\end{figure}

\textbf{\textit{Topological line node and surface states in CaAgP.--}}
%
%
%
Figure \ref{fig2}(a) shows the angle-resolved density of states, calculated from the surface Green's function via the QZ decomposition,\cite{miyata13, miyata15} on the (0001) surface terminated at a Ca$_3$P layer. 
Note that two types of termination, Ca$_3X$ and Ag$_3 X_2$, are possible on the (0001) surface of CaAg$X$. 
Hereafter, we focus only on the former type of termination.\cite{termination}
%
%
%
%
One can clearly see that a line node is located around the $\bar\Gamma$ point, which is projected from the bulk line node, and there exist surface states within the node.

The bulk line node is protected by  mirror-reflection symmetry with respect to the horizontal plane: $H(k_x, k_y, k_z) = M^\dag H(k_x, k_y, -k_z) M$.
The conduction and valence bands belong to $A_{2}''$ and $A_1'$ representations, respectively; therefore, these bands are degenerate on the mirror-reflection invariant plane of $k_z=0$.
%
Mirror-reflection symmetry allows one to introduce the $\mathbb Z_2$ topological invariant $\nu$ to characterize the band inversion as 
\begin{align}
  (-1)^{\nu(k_x, k_y)} = \xi(k_x, k_y, 0) \xi(k_x, k_y, \pi).
  \label{xi}
\end{align}
Here, $\xi(\bm k)$ is the product of eigenvalues of the mirror reflection for all the occupied bands at $\bm k$.
Obviously, $\nu(k_x, k_y)=1$ indicates band inversion from $k_z=0$ to $k_z=\pi$.
Moreover, it is worth mentioning that $\nu(k_x, k_y)$ is related to the Berry phase:\cite{Z2line} 
\begin{align}
 (-1)^{\nu(k_x, k_y)}  
 = \exp\left[
        \mathrm i 
	\int_{-\pi}^\pi \mathrm d k_z
        \mathrm{tr} A_z(\bm k)
        + \mathrm{tr} \ln B(k_x, k_y)
 \right],
 \label{bz2}
\end{align}
where 
the non-Abelian Berry connection is defined by
\begin{align}
 [\bm A(\bm k)]_{m n} = -\mathrm i \langle \bm k, m | \frac{\partial}{\partial \bm k} | \bm k, n \rangle,
\end{align}
where $|\bm k, m \rangle$ denotes an occupied eigenstate. 
$B(k_x, k_y)$ is the sewing matrix defined by
\begin{align}
 [B(k_x, k_y)]_{m n} = \langle (k_x, k_y, \pi), m | B_z | (k_x, k_y, \pi) - \bm G_z, n \rangle,
\end{align}
where $B_j$ is an operator satisfying
\begin{align}
 H(\bm k) = B^\dag_j H(\bm k + \bm G_j) B_j,
\end{align}
and $\bm G_j = 2\pi \hat{\bm x}_j$ denotes the $j$-th reciprocal lattice vector.
The calculated $\nu(k_x, k_y)$ is shown in Fig. \ref{fig2}(b).
The topological invariant is obtained to be nontrivial, $\nu(k_x, k_y)=1,$ within the line node, while trivial, $\nu(k_x, k_y)=0,$ in the outside.
Since $\nu(k_x, k_y)$ is invariant for a continuous change of parameters of the Hamiltonian, the line node is a topologically stable object. 

\begin{figure}
\centering
\ifnum \pdf=1
\includegraphics{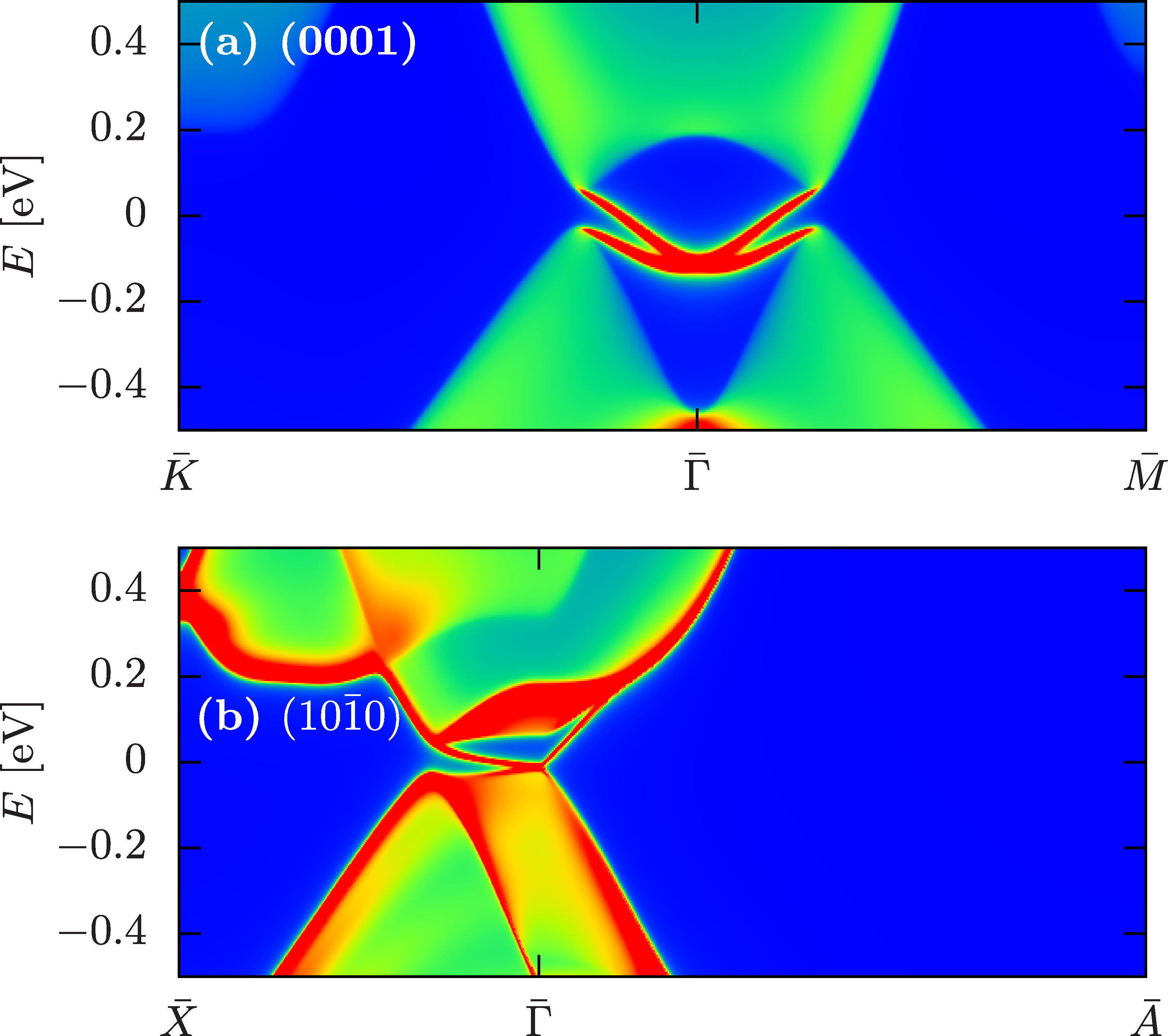}
\else
\includegraphics{fig3v2_outline.eps}
\fi
\caption{Angle-resolved density of states on the (a) Ca$_3 \mathrm{As}$-terminated (0001) and (b) $(10\bar 10)$ surfaces of CaAgAs.}
\label{fig3}
\end{figure}

\textbf{\textit{Topological insulating phase in CaAgAs.--}}
%
%
%
The degeneracy in dihedral point-group symmetry is lifted by the spin-orbit interaction.
Correspondingly,
the product of the eigenvalues of mirror reflection is always  unity: $\xi(k_x, k_y, 0) \xi(k_x, k_y, \pi) =1$; namely, the $\mathbb Z_2$ invariant takes the trivial value $\nu(k_x, k_y)=0$.\cite{absence} 
This means that the spin-orbit interaction yields an energy gap in the line node.
Furthermore, the spin-orbit interaction gives rise to a transition from a line-node semimetal to a topological insulator. 
Figure \ref{fig3} shows the angle-resolved density of states on the (0001) and $(10 \bar 10)$ surfaces. 
In the spin-orbit gap ($\sim$0.1 eV), gapless surface states appear. 
The Dirac point is located at the $\bar \Gamma$ point. 
On the (0001) surface [Fig. \ref{fig3}(a)], the gapless surface states are smoothly deformed from those in the absence of spin-orbit interaction [Fig. \ref{fig2}(a)].
On the $(10 \bar 10)$ surface, 
gapless surface states with approximately linear dispersion emerge in the induced band gap. 
The in-plane ($\bar \Gamma  \bar X$) velocity of the surface Dirac fermion is slower than the out-of-plane ($\bar \Gamma \bar A$) one, since the lattice constant along the $c$ axis is so small ($c/a \sim 0.6 $) that the interlayer coupling is much stronger.

The obtained gapless surface states, shown in Fig. \ref{fig3}, 
coincide with the $\mathbb Z_2$ topological invariants of $\nu_0;\nu_1\nu_2\nu_3=1;000$.
The simplified formula for the invariant derived by Fu and Kane\cite{fu07} cannot be applied to the present system owing to the lack of spatial-inversion symmetry.
Instead,
with the help of mirror-reflection symmetries with respect to the horizontal $H(k_x, k_y, k_z) = M_z^\dag H(k_x, k_y, k_z) M_z$ and the vertical $H(k_x, k_y, k_z) = M_y^\dag H(k_x, -k_y, k_z) M_y$ planes, 
the topological invariants can be easily calculated as 
\begin{align}
 \nu_{i \eta} = \frac{\Phi_{j l i \eta}(k_l=0) + \Phi_{j l i \eta}(k_l=\pi)}{2\pi} \mod 2,  
 \label{nu_mirror}
\end{align}
for $i,j,l=1,2,3$ ($i \ne j \ne l \ne i$) and $\eta=0, 1$.\cite{Z2mirror} 
The $\mathbb Z_2$ topological invariants are obtained by $\nu_0 = \nu_{i0} + \nu_{i1} \mod 2$, $\nu_{i} = \nu_{i1}$.
The Berry phase is given by
\begin{align}
 \Phi_{j l i \eta}(k_l) = \int_{-\pi}^\pi \mathrm d k_j \mathrm{tr} A_j(\bm k)|_{k_i= \eta \pi} - \mathrm i \, \mathrm{tr} \ln B_{j l i \eta}(k_l),
 \label{Phijki}
\end{align}
where
$B_{jli\eta}(k_l)$ is the sewing matrix defined by
\begin{align}
 [B_{j l i \eta}(k_l)]_{m n} = \langle \bm k, m | B_j | \bm k - \bm G_j, n \rangle|_{k_j = \pi, k_i=\eta \pi}.
\end{align}
Formula (\ref{Phijki}) is useful because only two points $k_l=0$ and $k_l=\pi$ are needed for the calculation, as in the case of spatial-inversion invariant systems.\cite{kim15}
%
The calculated Berry phases are summarized in Table \ref{berry}.
\begin{table}
\centering
\caption{Berry phase $\Phi_{jli\eta}(k_l)$ and $\mathbb Z_2$ invariant $\nu_{i \eta}$ of CaAgAs.}
\begin{tabular}{c|cc|cc|cc}
 \hline\hline
$i$ & 1 & 1 & 2 & 2 & 3 & 3
 \\
 $\eta$ & 0 & 1 & 0 & 1 & 0 & 1
  \\
  \hline
 $\Phi_{jli\eta}(k_l=0)$ & $2\pi$ & 0 & $2\pi$ & 0 & $2\pi$ & 0
  \\
  $\Phi_{jli\eta}(k_l=\pi)$ & 0 & 0 & 0 & 0 & 0 & 0
  \\
  \hline
 $\nu_{i \eta}$ & 1 & 0 & 1 & 0 & 1 & 0
  \\
  \hline\hline
\end{tabular}
\label{berry}
\end{table}
All the Berry phases on the zone boundaries $k_i=\pi$ are zero,
resulting in the topological invariants of $1;000$.\cite{1000}

\textbf{\textit{Effective model.--}}
%
%
%
We now derive effective models for CaAgP and CaAgAs.
In the model of CaAgAs (CaAgP), we (do not) take into account the spin-orbit interaction.

As discussed above, 
the conduction and valence bands at the $\Gamma$ point of CaAgP mainly consist of the $A_2''$ and $A_1'$ states, respectively.
Thus, the low-energy electronic structure is effectively described by the two-band model
\begin{align}
 H_{\rm CaAgP}(\bm k)
 = c(\bm k) + m(\bm k) \sigma_z + v k_z \sigma_y,
\end{align}
with $c(\bm k) = c_0 + c_1 k_z^2 + c_2 (k_x^2 + k_y^2)$ and $m(\bm k) = m_0 + m_1 k_z^2 + m_2 (k_x^2 + k_y^2)$ in the vicinity of the $\Gamma$ point.
A line node appears on the $k_z=0$ plane under the band-inversion condition of $m_0 m_2 < 0$.

In CaAgAs, the four-fold degenerated valence-band states at the $\Gamma$ point without the spin-orbit interaction, which consist of 
$E'$ states, are split into two doublets ($E_{3/2}$ and $E_{5/2}$) by introducing the spin-orbit interaction.
The resulting low-energy states consist of the $E_{5/2}$ doublet in the valence band and the $E_{3/2}$ doublet in the conduction band, which are described by the Hamiltonian
\begin{align}
 H_{\rm CaAgAs} (\bm k)
 = \begin{pmatrix}
        h(\bm k) & \Lambda(\bm k)
        \\
        \Lambda(\bm k)^\dag & h(-\bm k)^*
 \end{pmatrix}.
\end{align}
$h(\bm k)$ is the $2 \times 2$ Hamiltonian matrix for the states in which the $z$ components of the total angular momentum are given by $j_z = 1/2$ and by $j_z=3/2$. 
$h(\bm k)$ reads
\begin{align}
 h(\bm k) = c(\bm k) + m(\bm k) \sigma_z + A k_z (k_x \sigma_x - k_y \sigma_y),
\end{align}
up to the second order of $k$.
$\Lambda(\bm k)$ is the $2 \times 2$ matrix representing the spin-mixing effect induced by the spin-orbit interaction, defined by $\Lambda(\bm k) = \Lambda_1(\bm k) + \Lambda_2(\bm k)$, where
\begin{align}
 \Lambda_1(\bm k) &= -\mathrm i B_\perp k_z (1-\sigma_z) + -\mathrm i B_\parallel (k_x + \mathrm i k_y)\sigma_x,
 \\
 \Lambda_2(\bm k) &= D \sigma_y 
 \left[
        -\mathrm i \left(k_x^2-k_y^2 \right)  -2 k_x k_y
 \right].
\end{align}
$\Lambda(\bm k)$ yields a finite energy gap on the $k_z=0$ plane, thereby turning the system  into a strong topological insulator with 1;000.

These two effective Hamiltonians qualitatively describe the electronic structures of CaAgP and CaAgAs.
Therefore, it is useful for a further analysis of the electromagnetic responses and transport properties of these materials.
These Hamiltonians, however, do not quantitatively reproduce the low-energy electronic states, because the line node is located rather far from the $\Gamma$ point.
To obtain a quantitatively good model, one has to take into account higher energy states in addition to the low-lying states.

\textbf{\textit{Summary.--}}
We have clarified the topological electronic structure of hexagonal pnictides CaAg$X$ ($X= \mathrm P$ and As). 
CaAg$X$ is a line-node Dirac semimetal in the absence of spin-orbit interaction.
In reality,
CaAgP exhibits line-node-semimetal properties except in the very low energy and low-temperature regime owing to the tiny spin-orbit interaction ($\sim$10 K).
When P atoms are replaced with heavier As atoms, 
the strong spin-orbit interaction
widens the size of the band gap considerably at the line node.
 CaAgAs is found to be a strong topological insulator with the $\mathbb Z_2$ invariant of 1;000.

It has been known that the presence of reflection symmetry and band inversion is necessary and sufficient for the existence of a line node, whereas most of the line-node Dirac semimetals proposed so far preserve  spatial-inversion symmetry.
Moreover, the $\mathbb Z_2$ topological invariant characterizing the line node was defined in previous works in terms of $PT$ symmetry.\cite{fang15, kim-kane}
This invariant cannot be directly applied to  systems lacking inversion symmetry, including CaAg$X$ reported here. 
We have introduced
the alternative $\mathbb Z_2$ invariant $\nu(k_x, k_y)$ in this study, which is applicable to  systems without spatial-inversion symmetry.
This implies that line-node Dirac semimetals with and without spatial-inversion symmetry might exhibit intrinsically different electromagnetic responses and transport phenomena, which are related to the topological invariants.
Superconductivity of line-node semimetals is another interesting topic, as discussed in point-node Weyl\cite{cho12, shivamoggi13, wei14, wei14b, lu15, bendnik15} and Dirac semimetals.\cite{kobayashi15}
These issues will be addressed in future works.

\textbf{\textit{Note added.--}}
After the submission of this paper, we became aware of a
recent preprint\cite{chan15} in which the relation between the Berry phase and
eigenvalues of mirror reflection is discussed in a form different from
Eqs. (1) and (2).

\begin{acknowledgment}


The authors are grateful to S. Kobayashi for valuable discussion.

\end{acknowledgment}

\bibliographystyle{jpsj}
\bibliography{caagx}

\begin{thebibliography}{10}

\bibitem{thouless82}
D.~J. Thouless, M.~Kohmoto, M.~P. Nightingale, and M.~den Nijs: Phys. Rev.
  Lett. {\bfseries 49} (1982) 405.

\bibitem{kohmoto85}
M.~Kohmoto: Ann, Phys. {\bfseries 160} (1985) 343 .

\bibitem{hasan10}
M.~Z. Hasan and C.~L. Kane: Rev. Mod. Phys. {\bfseries 82} (2010) 3045.

\bibitem{qi11}
X.-L. Qi and S.-C. Zhang: Rev. Mod. Phys. {\bfseries 83} (2011) 1057.

\bibitem{tanaka12}
Y.~{Tanaka}, M.~{Sato}, and N.~{Nagaosa}: J. Phys. Soc. Jpn. {\bfseries 81}
  (2012) 011013.

\bibitem{murakami07}
S.~{Murakami}: New J. Phys. {\bfseries 9} (2007) 356.

\bibitem{wan11}
X.~Wan, A.~M. Turner, A.~Vishwanath, and S.~Y. Savrasov: Phys. Rev. B
  {\bfseries 83} (2011) 205101.

\bibitem{young12}
S.~M. Young, S.~Zaheer, J.~C.~Y. Teo, C.~L. Kane, E.~J. Mele, and A.~M. Rappe:
  Phys. Rev. Lett. {\bfseries 108} (2012) 140405.

\bibitem{steinberg14}
J.~A. Steinberg, S.~M. Young, S.~Zaheer, C.~L. Kane, E.~J. Mele, and A.~M.
  Rappe: Phys. Rev. Lett. {\bfseries 112} (2014) 036403.

\bibitem{fukushima08}
K.~Fukushima, D.~E. Kharzeev, and H.~J. Warringa: Phys. Rev. D {\bfseries 78}
  (2008) 074033.

\bibitem{zyuzin12}
A.~A. Zyuzin, S.~Wu, and A.~A. Burkov: Phys. Rev. B {\bfseries 85} (2012)
  165110.

\bibitem{hosur13}
P.~{Hosur} and X.~{Qi}: Compt. Rend. Phys. {\bfseries 14} (2013) 857.

\bibitem{burkov11}
A.~A. Burkov, M.~D. Hook, and L.~Balents: Phys. Rev. B {\bfseries 84} (2011)
  235126.

\bibitem{chiu14}
C.-K. Chiu and A.~P. Schnyder: Phys. Rev. B {\bfseries 90} (2014) 205136.

\bibitem{fang15}
C.~Fang, Y.~Chen, H.-Y. Kee, and L.~Fu: Phys. Rev. B {\bfseries 92} (2015)
  081201.

\bibitem{gao}
Z.~{Gao}, M.~{Hua}, H.~{Zhang}, and X.~{Zhang}: arXiv:1507.07504 .

\bibitem{mikitik06}
G.~P. Mikitik and Y.~V. Sharlai: Phys. Rev. B {\bfseries 73} (2006) 235112.

\bibitem{heikkila11}
T.~T. {Heikkil{\"a}} and G.~E. {Volovik}: JETP Lett. {\bfseries 93} (2011) 59.

\bibitem{heikkila}
T.~T. {Heikkila} and G.~E. {Volovik}: arXiv:1505.03277 .

\bibitem{phillips14}
M.~Phillips and V.~Aji: Phys. Rev. B {\bfseries 90} (2014) 115111.

\bibitem{mullen15}
K.~Mullen, B.~Uchoa, and D.~T. Glatzhofer: Phys. Rev. Lett. {\bfseries 115}
  (2015) 026403.

\bibitem{weng-bernevig}
H.~Weng, C.~Fang, Z.~Fang, B.~A. Bernevig, and X.~Dai: Phys. Rev. X {\bfseries
  5} (2015) 011029.

\bibitem{weng15}
H.~Weng, Y.~Liang, Q.~Xu, R.~Yu, Z.~Fang, X.~Dai, and Y.~Kawazoe: Phys. Rev. B
  {\bfseries 92} (2015) 045108.

\bibitem{chen}
Y.~{Chen}, Y.~{Xie}, S.~A. {Yang}, H.~{Pan}, F.~{Zhang}, M.~L. {Cohen}, and
  S.~{Zhang}: arXiv:1505.02284 .

\bibitem{kim-kane}
Y.~Kim, B.~J. Wieder, C.~L. Kane, and A.~M. Rappe: Phys. Rev. Lett. {\bfseries
  115} (2015) 036806.

\bibitem{yu15}
R.~Yu, H.~Weng, Z.~Fang, X.~Dai, and X.~Hu: Phys. Rev. Lett. {\bfseries 115}
  (2015) 036807.

\bibitem{zeng}
M.~{Zeng}, C.~{Fang}, G.~{Chang}, Y.-A. {Chen}, T.~{Hsieh}, A.~{Bansil},
  H.~{Lin}, and L.~{Fu}: arXiv:1504.03492 .

\bibitem{carter12}
J.-M. Carter, V.~V. Shankar, M.~A. Zeb, and H.-Y. Kee: Phys. Rev. B {\bfseries
  85} (2012) 115105.

\bibitem{chen15}
Y.~{Chen}, Y.-M. {Lu}, and H.-Y. {Kee}: Nature Communications {\bfseries 6}
  (2015) 6593.

\bibitem{kim15}
H.-S. Kim, Y.~Chen, and H.-Y. Kee: Phys. Rev. B {\bfseries 91} (2015) 235103.

\bibitem{liu}
J.~{Liu}, D.~{Kriegner}, L.~{Horak}, D.~{Puggioni}, C.~{Rayan Serrao},
  R.~{Chen}, D.~{Yi}, C.~{Frontera}, V.~{Holy}, A.~{Vishwanath}, J.~M.
  {Rondinelli}, X.~{Marti}, and R.~{Ramesh}: arXiv:1506.03559 .

\bibitem{rhim15}
J.-W. Rhim and Y.~B. Kim: Phys. Rev. B {\bfseries 92} (2015) 045126.

\bibitem{huh2}
Y.~{Huh}, E.-G. {Moon}, and Y.~B. {Kim}: arXiv:1506.05105 .

\bibitem{mitchell}
A.~K. {Mitchell} and L.~{Fritz}: arXiv:1506.05491 .

\bibitem{ramamurthy}
S.~T. {Ramamurthy} and T.~L. {Hughes}: arXiv:1508.01205 .

\bibitem{xie15}
L.~S. Xie, L.~M. Schoop, E.~M. Seibel, Q.~D. Gibson, W.~Xie, and R.~J. Cava:
  APL Mater. {\bfseries 3} (2015) 083602.

\bibitem{bian}
G.~{Bian}, T.-R. {Chang}, R.~{Sankar}, S.-Y. {Xu}, H.~{Zheng}, T.~{Neupert},
  C.-K. {Chiu}, S.-M. {Huang}, G.~{Chang}, I.~{Belopolski}, D.~S. {Sanchez},
  M.~{Neupane}, N.~{Alidoust}, C.~{Liu}, B.~{Wang}, C.-C. {Lee}, H.-T. {Jeng},
  A.~{Bansil}, F.~{Chou}, H.~{Lin}, and M.~{Zahid Hasan}: arXiv:1505.03069 .

\bibitem{schoop}
L.~M. {Schoop}, M.~N. {Ali}, C.~{Stra{\ss}er}, V.~{Duppel}, S.~S.~P. {Parkin},
  B.~V. {Lotsch}, and C.~R. {Ast}: arXiv:1509.00861 .

\bibitem{lu12}
L.~{Lu}, L.~{Fu}, J.~D. {Joannopoulos}, and M.~{Solja{\v c}i{\'c}}: Nat. Photo.
  {\bfseries 7} (2013) 294.

\bibitem{natori}
W.~M.~H. {Natori}, E.~{Miranda}, and R.~G. {Pereira}: arXiv:1505.06171 .

\bibitem{mewis79}
A.~Mewis: Zeitschrift f{\"u}r Naturforschung B {\bfseries 34} (1979) 14.

\bibitem{ishikawa14}
H.~Ishikawa, T.~Okubo, Y.~Okamoto, and Z.~Hiroi: J. Phys. Soc. Jpn. {\bfseries
  83} (2014) 043703.

\bibitem{WIEN2k}
P.~Blaha, K.~Schwarz, G.~Madsen, D.~Kvasnicka, and J.~Luitz: {\em {\it WIEN2k},
  An Augmented Plane Wave + Local Orbitals Program for Calculating Crystal
  Properties} (Techn. Universit\"ut Wien, Austria, 2001).

\bibitem{DFT}
We used the full-potential linearized augmented plane-wave method within the
  generalized gradient approximation as implemented in the WIEN2k
  code.\cite{WIEN2k} $24 \times 24 \times 36$ $k$-points sampling was used for
  the self-consistent calculation.

\bibitem{marzari97}
N.~Marzari and D.~Vanderbilt: Phys. Rev. B {\bfseries 56} (1997) 12847.

\bibitem{souza01}
I.~Souza, N.~Marzari, and D.~Vanderbilt: Phys. Rev. B {\bfseries 65} (2001)
  035109.

\bibitem{dorbital}
See Sec. S 4 of Supplemental Material.

\bibitem{miyata13}
T.~Miyata, S.~Honda, R.~Naito, and S.-L. Zhang: Jpn. J. Ind. Appl. Math.
  {\bfseries 30} (2013) 653.

\bibitem{miyata15}
T.~Miyata, R.~Naito, and S.~Honda: J. Eng. Math.  (2015) 1.

\bibitem{termination}
Electronic states on the Ag$_3X_2$ termination is shown in Sec. S 1 of
  Supplemental Material.

\bibitem{Z2line}
Equation (\ref{xi}) is derived from Eq. (\ref{bz2}). See Sec. S 2 of
  Supplemental Material.

\bibitem{absence}
See Sec. S 2.4 of Supplemental Material.

\bibitem{fu07}
L.~Fu and C.~L. Kane: Phys. Rev. B {\bfseries 76} (2007) 045302.

\bibitem{Z2mirror}
Equation (\ref{nu_mirror}) is obtained for mirror-reflection-invariant systems.
  See Sec. S 3 of Supplemental Material.

\bibitem{1000}
Dirac semimetals in which a line node is located around the $\Gamma$ point
  always becomes a strong topological insulator with $\nu_0;\nu_1\nu_2\nu_3 =
  1;000$ when spin-orbit interaction is switched on. See Sec. S 3.3 of
  Supplemental Material.

\bibitem{cho12}
G.~Y. Cho, J.~H. Bardarson, Y.-M. Lu, and J.~E. Moore: Phys. Rev. B {\bfseries
  86} (2012) 214514.

\bibitem{shivamoggi13}
V.~Shivamoggi and M.~J. Gilbert: Phys. Rev. B {\bfseries 88} (2013) 134504.

\bibitem{wei14}
H.~Wei, S.-P. Chao, and V.~Aji: Phys. Rev. B {\bfseries 89} (2014) 014506.

\bibitem{wei14b}
H.~Wei, S.-P. Chao, and V.~Aji: Phys. Rev. B {\bfseries 89} (2014) 235109.

\bibitem{lu15}
B.~Lu, K.~Yada, M.~Sato, and Y.~Tanaka: Phys. Rev. Lett. {\bfseries 114} (2015)
  096804.

\bibitem{bendnik15}
G.~Bednik, A.~A. Zyuzin, and A.~A. Burkov: Phys. Rev. B {\bfseries 92} (2015)
  035153.

\bibitem{kobayashi15}
S.~{Kobayashi} and M.~{Sato}: arXiv:1504.07408 .

\bibitem{chan15}
Y.-H. {Chan}, C.-K. {Chiu}, M.~Y. {Chou}, and A.~P. {Schnyder}:
  arXiv:1510.02759 .

\bibitem{kobayashi2}
S. Kobayashi \textit{et al.}, in preparation.

\bibitem{yu11}
R.~Yu, X.~L. Qi, A.~Bernevig, Z.~Fang, and X.~Dai: Phys. Rev. B {\bfseries 84}
  (2011) 075119.

\bibitem{fu10}
L.~Fu and E.~Berg: Phys. Rev. Lett. {\bfseries 105} (2010) 097001.

\bibitem{qi10}
X.-L. Qi, T.~L. Hughes, and S.-C. Zhang: Phys. Rev. B {\bfseries 81} (2010)
  134508.

\bibitem{sato10}
M.~Sato: Phys. Rev. B {\bfseries 81} (2010) 220504.

\end{thebibliography}

\def\thesection{S \arabic{section}}
\def\theequation{S\arabic{equation}}
\renewcommand{\thefigure}{S \arabic{figure}}
 \setcounter{figure}{0}
\setcounter{section}{0}

\section{Termination-dependent (0001) surface states }

CaAg$X$ consists of alternating stacking of Ca$_3X$ and Ag$_3X_2$ layers along the $c$ axis, i.e.,
the (0001) surface can be terminated by either of these layers.
Correspondingly,
energy dispersion relations of (0001) surface states on the Ca$_3X$ termination are different from those on the Ag$_3X_2$ termination.
\begin{figure}
 \centering
\ifnum \pdf=1
\includegraphics{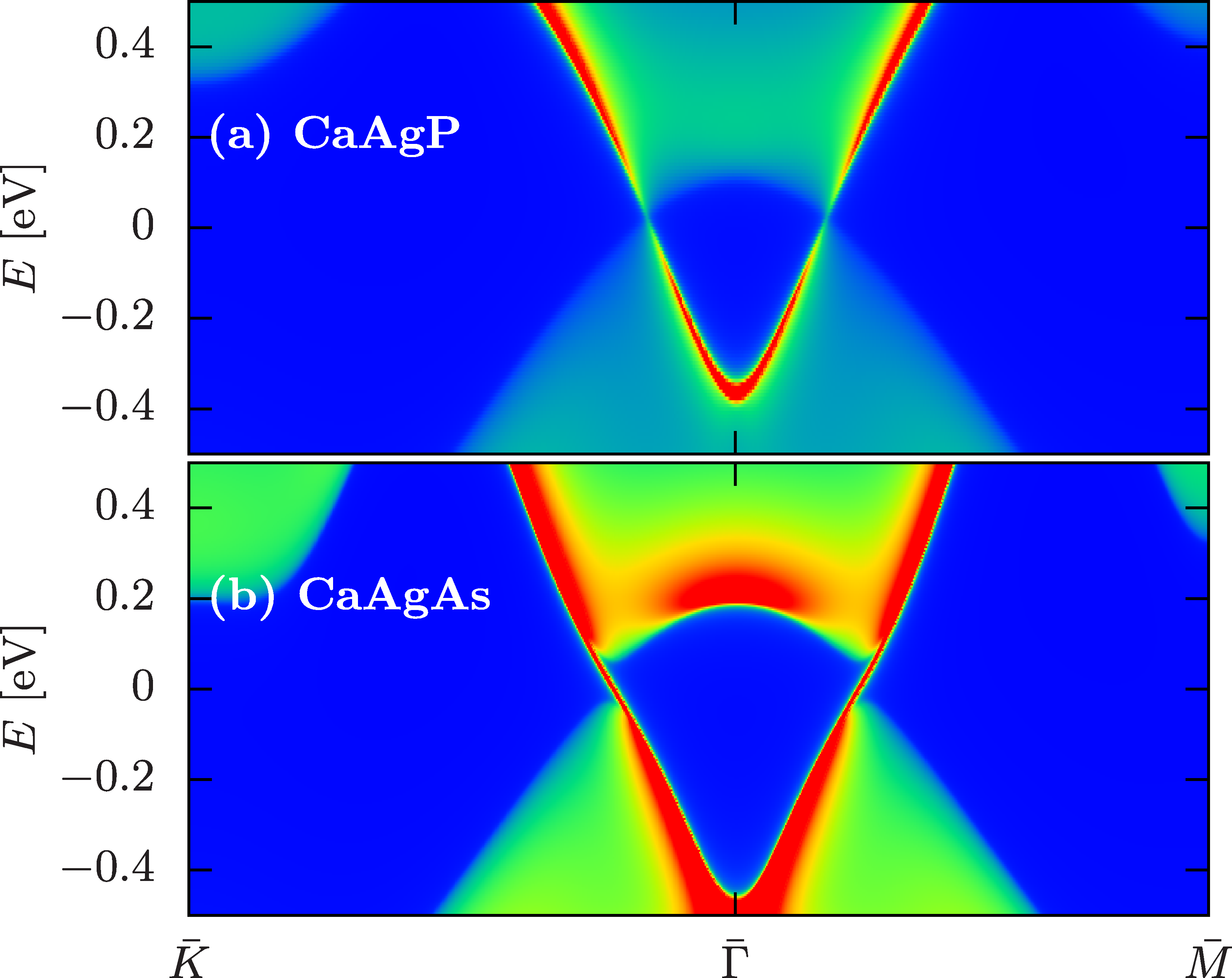}
\else
\includegraphics{top-bottom_outline.eps}
\fi
\caption{Angle-resolved density of states on the (0001) surface with the Ag$_3X_2$ termination for $X= \mathrm P$ (a),  As (b).}
\label{top-bottom}
\end{figure}
Angle-resolve density of states on the Ag$_3X_2$ termination is shown in Fig. \ref{top-bottom}.
Surface states on the Ca$_3X$ termination of CaAgP (CaAgAs) show up within the (gapped) line node, as shown in Fig. 2(a) [3(a)], while those on the Ag$_3X_2$ termination do not, as shown in Fig. \ref{top-bottom}.

\section{$\mathbb Z_2$ line node protected by mirror-reflection symmetry}

A line node protected by mirror-reflection symmetry is simply understood as follows: conduction and valence bands have opposite parity of the reflection hence these are degenerated at the reflection invariant plane $k_z=0$ or $k_z=\pi$.
Here, we show that the topological invariant which is associated with a line node can be defined in terms of Berry phase. 

\subsection{Definition}
Suppose a system preserving mirror-reflection symmetry with respect to the horizontal plane.
The Hamiltonian $H(\bm k)$ satisfies
\begin{align}
  H(k_x, k_y, k_z) = M^\dag H(k_x, k_y, -k_z) M.
\end{align}
Mirror-reflection operator $M$, in general, depends on $k_z$ in systems with sublattice structure.
In the following, we choose the gauge in which $M$ is independent of $k_z$.
Instead, the Hamiltonian has a nontrivial periodicity:
\begin{align}
 H(\bm k) = B^\dag H(\bm k + \bm G_z) B,
\end{align}
where $\bm G_z = (0,0,2\pi)$ denotes the reciprocal lattice vector along the $z$ axis.
The $\mathbb Z_2$ topological invariant $\nu(k_x, k_y)$ is defined by
\begin{align}
 (-1)^{\nu(k_x, k_y)} = \exp\left[ \mathrm i \int_{-\pi}^\pi {\mathrm d k_z} \mathrm{tr} A_z(\bm k) + \mathrm{tr} \ln B(k_x, k_y) 
\right].
\label{nu}
\end{align}
The non-Abelian Berry connection $\bm A(\bm k)$ and unitary matrix $B(k_x, k_y)$ are defined by
\begin{align}
 [\bm A(\bm k)]_{m n} = -\mathrm i \langle \bm k, m | \frac{\partial}{\partial \bm k} | \bm k, n \rangle,
\end{align}
and
\begin{align}
 [B(k_x, k_y)]_{m n} = \langle (k_x, k_y, \pi), m | B | (k_x, k_y, -\pi), n \rangle.
\end{align}

\subsection{Gauge symmetry}

Next, we show that
$\nu(k_x, k_y) \mod 2$ is invariant under a U($N$) gauge transformation $|\bm k, m\rangle \to |\bm k, n \rangle [g(\bm k)]_{nm}$, $g(\bm k) \in \mathrm U(N)$.
$\bm A(\bm k)$ and $B(k_x, k_y)$ are transformed into
\begin{align}
 \bm A(\bm k) &\to g^\dag(\bm k) \bm A(\bm k) g(\bm k) - \mathrm i g^\dag(\bm k) \frac{\partial g(\bm k)}{\partial \bm k},
 \\
 B(k_x, k_y) & \to
 g^\dag(k_x, k_y, \pi) B(k_x, k_y) g(k_x, k_y, -\pi).
\end{align}
From the integral
\begin{align}
 \int_{-\pi}^\pi \mathrm d k_z \mathrm{tr} g^\dag(\bm k) \frac{\partial g(\bm k)}{\partial k_z}
  = \ln \frac{\det g(k_x, k_y, \pi)}{\det g(k_x, k_y, -\pi)} + \mathrm i  2 n \pi, 
  \ 
  n \in \mathbb Z,
\end{align}
one can verify that the right hand side of Eq. (\ref{nu}) is invariant under the gauge transformation.

\subsection{$\mathbb Z_2$ invariant and mirror-reflection symmetry}
\label{z2_mirror}

If the system has mirror-reflection symmetry, $\nu(k_x, k_y)$ is the $\mathbb Z_2$ invariant.
The mirror-reflection symmetry requires the following relation;
\begin{align}
 H(k_x, k_y, k_z) = M^\dag H(k_x, k_y, -k_z) M.
\end{align}
We introduce the unitary matrix $M(\bm k)$ in the occupied subspace;
\begin{align}
  [M(k_x, k_y, k_z)]_{m n} = \langle (k_x, k_y, k_z), m | M | (k_x, k_y, -k_z), n \rangle.
\end{align}
On the reflection invariant plane of $k_z=0$, the states are simultaneously the eigenstates of $M$ hence
\begin{align}
 \det M(k_x, k_y, 0) = \pm 1,
\end{align}
where the phase of $M$ is fixed as $M^2=1$.
On the other reflection invariant plane of $k_z=\pi$, the mirror-reflection operator is modified to
\begin{align}
 M' = M B^\dag.
\end{align}
$M'$ satisfies
\begin{align}
 [H(k_x, k_y, \pi), M'] = 0.
\end{align}
Unitary matrix defined by
\begin{align}
 [M'(k_x, k_y, \pi)]_{m n} = \langle (k_x, k_y, \pi), m | M' | (k_x, k_y, \pi), n \rangle,
\end{align}
also satisfies
\begin{align}
 \det M'(k_x, k_y, \pi) = \pm 1.
\end{align}

Now we prove that $\nu(k_x, k_y)$ is the $\mathbb Z_2$ invariant.
The non-Abelian Berry connection satisfies
\begin{align}
 A_z(k_x, k_y, k_z) &= -M^\dag(k_x, k_y, -k_z) A_z(k_x, k_y, -k_z) M(k_x, k_y, -k_z) 
\nonumber\\ & \quad
+ \mathrm i M^\dag(k_x, k_y, -k_z) \frac{\partial M(k_x, k_y, -k_z)}{-\partial k_z}.
\end{align}
Therefore, the integral of $\bm A$ reduces to
\begin{align}
 \int_{-\pi}^0 \mathrm d k_z \mathrm{tr} A_z(\bm k)
 = -\int_{0}^\pi \mathrm d k_z \mathrm{tr} A_z(\bm k)
 + \mathrm i \int_0^\pi \mathrm d k_z \mathrm{tr} M^\dag(\bm k) \frac{\partial M(\bm k)}{\partial k_z}.
\end{align}
The second term is rewritten as 
\begin{align}
   \mathrm i \int_0^\pi \mathrm d k_z \mathrm{tr} M^\dag(\bm k) \frac{\partial M(\bm k)}{\partial k_z}
   = \mathrm i \ln \frac{\det M(k_x, k_y, \pi)}{\det M(k_x, k_y, 0)} + 2 n \pi.
\end{align}
Consequently, one obtains
\begin{align}
 \mathrm i \int_{-\pi}^\pi \mathrm d k_z \mathrm{tr} A_z(\bm k)
 + \mathrm{tr} \ln B(k_x, k_y)
 = - \ln \frac{\det M'(k_x, k_y, \pi)}{\det M(k_x, k_y, 0)} + \mathrm i 2 n \pi,
 \label{Phi2}
\end{align}
namely
\begin{align}
 (-1)^{\nu(k_x, k_y)}  
 = \frac{\det M'(k_x, k_y, \pi)}{\det M(k_x, k_y, 0)} = \pm 1,
 \label{Phi3}
\end{align}
which is the same as Eq. (1) in the main manuscript.

\subsection{Topological invariant in spinful systems}
\label{absence}

In spinful systems, the topological invariants of spin up and down may cancel each other out; $\nu(k_x, k_y)=0$, owing to time-reversal $T$ symmetry.
The mirror reflection in spinful systems involves the spin hence $T^{-1} M T = -M$, and
\begin{align}
 M(\bm k)^* = - T(-\bm k)^\dag M(-\bm k) T(-\bm k),
\end{align}
where unitary skew matrix $T(\bm k)$ is defined by
\begin{align}
 [T(\bm k)]_{m n} = \langle \bm k, m | T \left| -\bm k, n \right\rangle.
\end{align}
As a result, one obtains
\begin{align}
 \mathrm{tr} M(k_x, k_y, 0) &= -\mathrm{tr} M(-k_x, -k_y, 0),
\\ 
 \mathrm{tr} M'(k_x, k_y, \pi) &= -\mathrm{tr} M'(-k_x, -k_y, \pi).
\end{align}
For $(k_x, k_y, \Gamma)$, $\Gamma = 0, \pi$, which is continuously connected (without gap closing) to a time-reversal invariant momentum $\bm \Gamma$ within the $k_z=\Gamma$ plane, the following relation holds
\begin{align}
 \mathrm{tr} M(k_x, k_y, 0) = \mathrm{tr} M'(k_x, k_y, \pi) = 0,
\end{align}
since $\mathrm{tr} M(\bm \Gamma) = \mathrm{tr} M'(\bm \Gamma) = 0$ and $\mathrm{tr} M(k_x, k_y, \Gamma)$ is a quantized invariant.
Consequently, the number of occupied states with the eigenvalue of $+\mathrm i$ of mirror reflection are the same as that with the eigenvalue of $-\mathrm i$.
This proves that $\det M(k_x, k_y, 0) = \det M(k_x, k_y, \Gamma) = (-1)^{N/2}$,  $\nu(k_x, k_y) = 0$, where $N$ denotes the number of occupied bands, and that there is no line node encircling time-reversal invariant momenta.
Note that line nodes not around time-reversal invariant momenta may appear in a case that antisymmetric spin-orbit interaction is much stronger than symmetric one,\cite{kobayashi2} nevertheless it is not the case in an actual material CaAg$X$.


\section{$\mathbb Z_2$ invariant and mirror-reflection symmetry in the presence of spin-orbit interaction}

We show that
the $\mathbb Z_2$ invariant characterizing insulators which respect both time-reversal and mirror-reflection symmetries reduces to the one-dimensional topological invariant Eq. (1).

\subsection{Definition}
We start with the following expression
\begin{align}
 \nu = \int_0^\pi \frac{\mathrm d k_y}{2\pi} \frac{\partial \Phi(k_y)}{\partial k_y} - \frac{\Phi(\pi) - \Phi(0)}{2\pi} \mod 2.
 \label{nu1}
\end{align}
$\Phi(k_y)$ is the Berry phase defined by
\begin{align}
 \Phi(k_y) = \int_{-\pi}^\pi \mathrm d k_x \mathrm{tr} A_x(\bm k) - \mathrm i \ \mathrm{tr} \ln B(k_y),
 \label{B}
\end{align}
with
\begin{align}
  [B(k_y)]_{m n} = \langle (\pi, k_y), m | B | (-\pi, k_y), n \rangle.
\end{align}
Operator $B$ satisfies 
\begin{align}
 H(k_x, k_y) = B^\dag H(k_x + 2\pi, k_y) B.
\end{align}
Equations (\ref{nu1}) and (\ref{B}) are equivalent to the $\mathbb Z_2$ invariant derived in Ref \cite{yu11}. 
This is obviously confirmed in a particular choice of gauge in which $B$ is the identity.

\subsection{Mirror-reflection symmetry and reduced formula}
Now we prove 
\begin{align}
  \nu = \frac{\Phi(0) + \Phi(\pi)}{2\pi} \mod 2,
  \label{nu2}
\end{align}
in the presence of mirror-reflection symmetry as
\begin{align}
 H(k_x, k_y) = M^\dag H(-k_x, k_y) M.
\end{align}
In a manner similar to Secs. \ref{z2_mirror} and \ref{absence}, $\Phi(k_y)$ reduces to
\begin{align}
 \Phi(k_y) = 2 n \pi, 
 \
 n \in \mathbb Z.
\end{align}
$\Phi(k_y) \mod 2 \pi$ for $k_y \ne 0, \pi$ is gauge invariant, while on the time-reversal invariant momenta $\Phi(0) \mod 4\pi$ and $\Phi(\pi) \mod 4\pi$ are.\cite{yu11}
This means that only $\Phi(0)$ and $\Phi(\pi)$ determines the $\mathbb Z_2$ invariant.
Then we arrive at Eq. (\ref{nu2}).

\subsection{Line node and strong topological insulator}

Suppose a system hosting a line node located on the $k_z=0$ plane around the $\Gamma$ point in the Brillouin zone, as in the case of CaAg$X$ discussed in the main manuscript.
Spin-orbit interaction induces an energy gap at the line node.
Here we calculate the $\mathbb Z_2$ topological invariant.
On the $k_x=0$ and $k_x=\pi$ planes, the invariant is given by
\begin{align}
 \nu_{1 \eta} = \frac{\Phi_{321 \eta}(0) + \Phi_{321 \eta}(\pi)}{2\pi} \mod 2,
\end{align}
where the subscript is defined in Eq. (7).
Berry phase $\Phi_{321 \eta}(\Gamma)$ is given by the integral along the $k_z$ axis on the zone center and zone boundary, where the energy gap does not vanish as one turns off the spin-orbit interaction.
From this fact, the Berry phase reduces to that in the absence of spin-orbit interaction;
\begin{align}
 \Phi_{321 \eta}(\Gamma) = \Phi_{321 \eta}^\uparrow (\Gamma) + \Phi_{321 \eta}^\downarrow (\Gamma)
  = 2 \Phi_{321 \eta}^\uparrow (\Gamma),
\end{align}
where $\Phi^\uparrow$ and $\Phi^\downarrow$ denote the Berry phase in the spin-up and spin-down subspaces without spin-orbit interaction.
A similar technique is found in Refs.\cite{fu10, qi10, sato10}
Furthermore, from Eqs. (\ref{Phi2}) and (\ref{Phi3}), 
\begin{align}
 \Phi_{321 \eta}(\Gamma) = 2 \pi \nu(\eta \pi, \Gamma) \mod 4 \pi.
\end{align}

The $\mathbb Z_2$ invariant $\nu(k_x, k_y)$ associated with a line node is obtained to be $\nu(k_x, k_y)=1$ [$\nu(k_x, k_y)=0$] within (out of) the line node.
Therefore, 
$\nu_{i \eta} = 1$ when the four time-reversal invariant momenta on the $k_i = \eta \pi$ plane are enclosed by an odd number of line nodes, otherwise $\nu_{i\eta} = 0$.
For instance, in the case that there is a single line node around $\bm \Gamma_{\rm line} = (\Gamma_{\rm line}^x, \Gamma_{\rm line}^y, \Gamma_{\rm line}^z)$, the topological invariants are obtained to be 
\begin{align}
(\nu_1, \nu_2) = 
\left(\Gamma_{\rm line}^x, \Gamma_{\rm line}^y
\right)/\pi, 
\
 \nu_0 = 1.
\end{align}
For the case of CaAg$X$, because a single line node appears around the $\Gamma$ point, one gets
\begin{align}
 (\nu_1, \nu_2) = (0, 0),
 \
 \nu_0 = 1.
\end{align}
\noindent
And also, one can calculate $\nu_3$ in a similar manner since an additional mirror-reflection symmetry with respect to the $x z$ plane is satisfied and the system has an energy gap and no integral pass along the $k_y$ direction through the line node on the $k_z=\pi$ plane in the absence of spin-orbit interaction,
In consequence, we obtain
\begin{align}
   \nu_{31} = 0.
\end{align}
The resultant $\mathbb Z_2$ invariant is given by 1;000.
%

\section{27-orbital model}

Here we show bulk and surface electronic states of a tight-binding model for CaAg$X$ which consists of the 27 orbitals, i.e., $d$ orbitals of Ca, $s$ orbitals of Ag, and $p$ orbitals of $X$.
\begin{figure}
\centering
\ifnum \pdf=1
\includegraphics{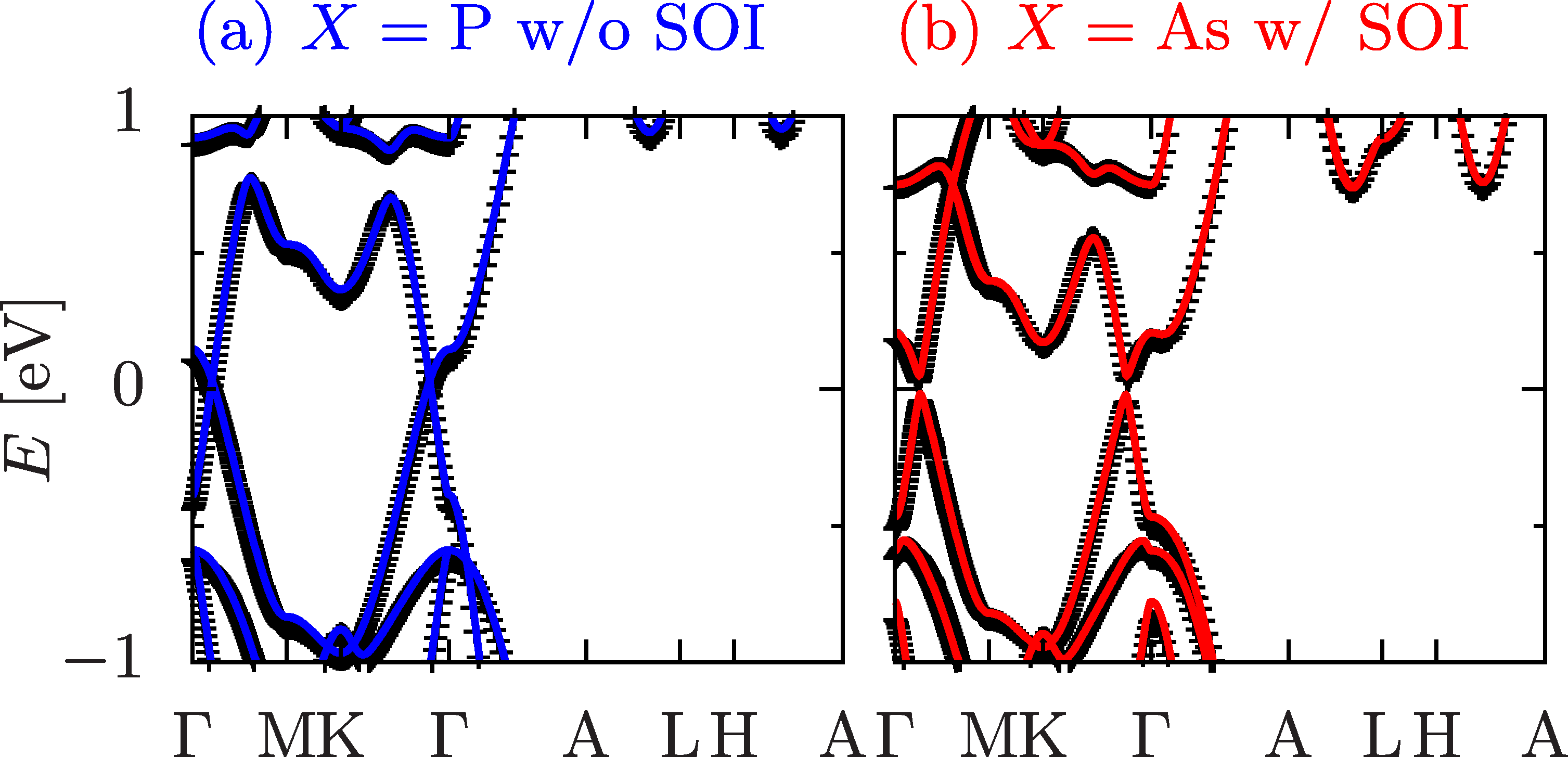}
\else
\includegraphics{band27.eps}
\fi
\caption{Bulk electronic states of (a) CaAgP without spin-orbit interaction and (b) CaAgAs with spin-orbit interaction.
Energy bands obtained by using the 27-orbital tight-binding model and by first-principles calculation are denoted by the solid lines and by the crosses, respectively.
}
\label{band27}
\end{figure}
Similarly to the 12-orbital model as discussed in the main context, spin-orbit interaction of CaAgP is neglected.

In CaAgAs, on the other hand, spin-orbit interaction only for the $p$ orbitals of As atoms is taken into account  in the form of $H_{\rm SOI} = \lambda \bm L \cdot \bm S$ with $\lambda = 0.07$ eV.
\begin{figure}
\centering
\ifnum \pdf=1
\includegraphics{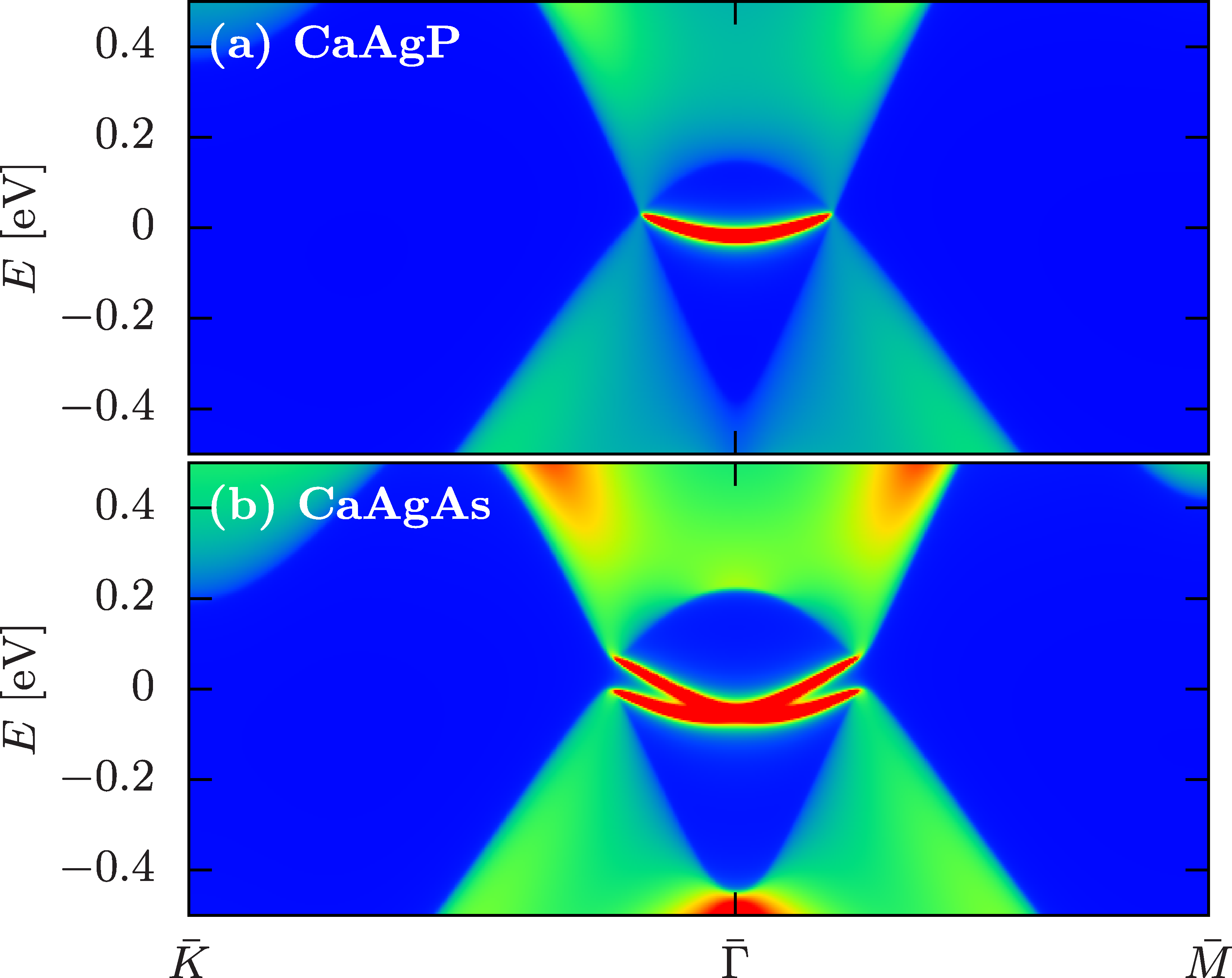}
\else
\includegraphics{27orbital.eps}
\fi
\caption{Angle-resolved density of states obtained by using the 27-orbital tight-biding model on the (0001) surface with the Ca$_3X$ termination for $X=$P (a), As (b).}
\label{27orbital}
\end{figure}
The bulk and surface electronic states are shown in Figs. \ref{band27} and \ref{27orbital}, respectively.
The obtained energy bands for both CaAgP 
[Fig. \ref{band27}(a)]
and CaAgAs [Fig. \ref{band27}(b)] well coincide with the first-principles bands not only in the low-energy regime ($E \sim 0$ eV) but also in high-energy regime ($E \sim 1$ eV).
The electronic states for CaAg$X$ on the (0001) surface with Ca$_3X$ termination are shown in Fig. \ref{27orbital}, which are qualitatively the same as those of the 12-orbital models shown in Figs. 2(a) and 3(a).
One can confirm that the model dependence of electronic states is negligibly small. 

\clearpage
\end{document}